\documentclass[prb,letterpaper,aps,floatfix,twocolumn]{revtex4-1}
\usepackage{graphicx}
\usepackage{amsmath}
\usepackage[small,bf]{subfigure}
\newcommand{\beq}{\begin{equation}}
\newcommand{\eeq}{\end{equation}}
\newcommand{\bk}{{{\bf{k}}}}

\newcommand{\bq}{{\bf{q}}}

\newcommand{\bb}{{\bf{b}}}

\newcommand{\be}{{\bf e}}
\newcommand{\beqa}{\begin{eqnarray}}
\newcommand{\eeqa}{\end{eqnarray}}

\newcommand{\bsigma}{{\boldsymbol \sigma}}
\newcommand{\btau}{{\boldsymbol \tau}}

\begin{document}
\title{Quantum anomalies in nodal line semimetals}
\author{A.A. Burkov}
\affiliation{Department of Physics and Astronomy, University of Waterloo, Waterloo, Ontario 
N2L 3G1, Canada} 
\date{\today}
\begin{abstract}
Topological semimetals is a new class of condensed matter systems with nontrivial electronic structure topology. 
Their unusual observable properties may often be understood in terms of quantum anomalies.
In particular, Weyl and Dirac semimetals, which have point band touching nodes, 
are characterized by the chiral anomaly, which leads to the Fermi arc surface states, anomalous Hall effect, negative longitudinal magnetoresistance and planar Hall effect. 
In this paper we explore analogous phenomena in nodal line semimetals. We demonstrate that such semimetals realize a three dimensional analog of the parity anomaly, which is a known property of two dimensional Dirac semimetals arising, for example, on the surface of a three dimensional topological insulator. 
We relate one of the characteristic properties of nodal line semimetals, namely the drumhead surface states, to this anomaly, and derive the field theory, which encodes the corresponding anomalous response. 
\end{abstract}
\maketitle
\section{Introduction}
\label{sec:1}
Topological (semi)metal is a new phase of matter, which is characterized by a nontrivial electronic structure topology, yet is not an insulator with a gap in the spectrum.~\cite{Weyl_RMP,Hasan_ARCMP,Felser_ARCMP,Burkov_ARCMP,Wan11,Burkov11-1,Burkov11-2,Xu11,Kane12,Fang12,Fang13,Chen14,Neupane14,HasanTaAs,DingTaAs2,DingTaAs,Lu15,Felser17}
Integer momentum space invariants, which characterize topologically nontrivial states of matter, are defined in this case on the Fermi surface, rather than in the whole Brillouin zone (BZ).~\cite{Volovik03,Haldane04,Volovik07}
Such Fermi surface invariants arise from singularities in the electronic structure, in the simplest case isolated points, at which 
different bands touch. 
The significance of such points was emphasized early on by Volovik,~\cite{Volovik03,Volovik07} and also pointed out by Murakami.~\cite{Murakami07} 

Nontrivial electronic structure topology has both spectroscopic manifestations, in the form of localized edge states, and manifestations
in response, in the form of quantized, or just insensitive to perturbations and microscopic details, transport or other response properties. 
In the context of Weyl and Dirac semimetals this topological response may have different manifestations, such as negative 
longitudinal magnetoresistance,~\cite{Spivak12,Burkov_lmr_prb,Ong_anomaly,Li_anomaly} giant planar Hall 
effect,~\cite{Burkov_gphe,Tewari_gphe,Shen_gphe,Felser_gphe,Zhang_gphe} and anomalous Hall effect.~\cite{Burkov_AHE,Felser17}

The edge states and the topological response are closely related and one way to understand this relation is in terms of the concept of quantum anomalies. 
Quantum anomaly is often described as violation of a classical symmetry (i.e. symmetry of the action) by quantum effects in the 
presence of external (say electromagnetic) fields.~\cite{Fujikawa79}
This violation of a symmetry then leads to nonconservation of a current, which should be conserved classically by Noether's theorem. 
This viewpoint, however, is strictly applicable only to quantum anomalies in the particle physics context, since condensed matter systems typically do not possess the corresponding symmetries, such as the chiral symmetry, to begin with. 

Another, more useful in the condensed matter context, way to understand the anomalies is in terms of failure of gauge invariance. 
Current nonconservation may be represented in terms of an anomalous term in the action for the electromagnetic (or some other) 
field, which is not gauge invariant in the presence of a boundary. 
What restores the overall gauge invariance is the contribution of edge states, which precisely cancels the gauge invariance violating part of the bulk action.~\cite{CallanHarvey}

It is instructive to see how this works in the case of the simplest topological metal system: a Weyl semimetal with two 
nodes.~\cite{Goswami_anomalies}
The chiral anomaly in this system may be expressed in terms of the following action for the electromagnetic field~\cite{Zyuzin12-1}
\beq
\label{eq:1}
S = - \frac{e^2}{4 \pi^2} \int d t \, \, d^3 r\, q_{\mu} \epsilon^{\mu \nu \alpha \beta} A_{\nu} \partial_{\alpha} A_{\beta}, 
\eeq
where $q_{\mu} = q \delta_{\mu z}$ describes the momentum space separation between the Weyl nodes ($2q$) along the 
$z$-axis and $\hbar = c =1$ units are used here and throughout this paper, except in some of the final formulas. 
We have also ignored the chiral magnetic effect~\cite{Kharzeev08} and related phenomena in Eq.~\eqref{eq:1} for 
simplicity (these effects exist only away from equilibrium and require certain modifications of Eq.~\eqref{eq:1}, which 
we do not want to discuss here; they have been discussed in detail, for example in Refs.~\onlinecite{Franz13,Chen13}). 

Consider a gauge transformation $A_{\mu} \rightarrow A_{\mu} + \partial_{\mu} \chi$. 
This changes the action in Eq.~\eqref{eq:1} by 
\beq
\label{eq:2}
S_{\chi} = - \frac{e^2}{4 \pi^2} \int d t \,\, d^3r\, q_{\mu} \epsilon^{\mu \nu \alpha \beta} \partial_{\nu} \chi \partial_{\alpha} A_{\beta}. 
\eeq
$S_{\chi}$ vanishes identically in the absence of boundaries (i.e. in a system with periodic boundary conditions), but does not vanish 
in a system with a boundary.
Indeed, suppose $q_{\mu} = q \delta_{\mu z} \Theta(x)$, where $\Theta(x)$ is the Heaviside step function. This may represent, for example, a contact between a Weyl semimetal in the $x > 0$ half-space and vacuum in the $x < 0$ half-space. 
Then we obtain
\beq
\label{eq:3}
S_{\chi} = \frac{e^2 q}{2 \pi^2} \int d t \, dy\, dz \,\chi (\partial_y A_0 - \partial_t A_y). 
\eeq
Eq.~\eqref{eq:3} clearly does not vanish in general and this means that Eq.~\eqref{eq:1} fails to be gauge invariant in the presence 
of a boundary. 
This failure of gauge invariance is a symptom that Eq.~\eqref{eq:1} is incomplete. 
Indeed, what it does not take into account is the Fermi arc edge states, which are present in the Weyl semimetal system, 
described by Eq.~\eqref{eq:1}. 
For a Weyl semimetal with two nodes, separated along the $z$-axis and with a boundary, perpendicular to the $x$-axis, these edge states have the form of a chiral sheet, which disperses along the $y$-direction and extends between the projections 
of the Weyl node locations onto the surface BZ. 
This chiral sheet may be viewed as  $2q/(2\pi/L_z) = q L_z/\pi$ (where $L_z$ is the size of the system in the $z$-direction) one dimensional chiral modes, which upon a gauge transformation generate a contribution to the action, which is equal to $-S_{\chi}$, thus cancelling the non-gauge-invariant part of the bulk 
action, as described in detail in Ref.~\onlinecite{Goswami_anomalies}.

In this paper we explore the connection between the edge states and the quantum anomalies in the context of nodal line 
semimetals.~\cite{Burkov11-2,Neupane_nodalline,Hasan_nodalline,Schoop_nodalline}
Nodal line semimetals possess surface states that have the form of a drumhead: the surface states are weakly dispersing and exist
in a two dimensional (2D) region in the crystal momentum space, bounded by the projection of the bulk nodal line onto the surface 
BZ.~\cite{Burkov11-2} 
A natural question one may ask is if there exists a bulk anomaly, such as the chiral anomaly described above, which is associated with the drumhead surface states? 

We answer this question in the affirmative and demonstrate that the quantum anomaly, associated with the drumhead surface states of nodal line semimetals is closely related to the parity anomaly of (2+1)-dimensional relativistic Dirac fermions.
Somewhat related ideas were put forward recently in Ref.~\onlinecite{Schnyder17},  
while an entirely different viewpoint was presented earlier in Ref.~\onlinecite{Hughes_linenode}. 
Closely related phenomena in Dirac semimetals in an external magnetic field have been discussed in Ref.~\onlinecite{Burkov18-1}. 

The rest of this paper is organized as follows. 
In Section~\ref{sec:2} we introduce a model of a thin three dimensional (3D) topological insulator (TI) 
film, doped with magnetic impurities, and analyze the behavior of this system as a function of the magnetization 
direction, in particular focusing on a quantum phase transition between two quantum anomalous Hall states of the film 
with opposite signs of the Hall conductivity. 
We demonstrate that at the critical point this system exhibits two massless Dirac fermions, separated in momentum space, and 
1D edge states, connecting the two Dirac fermions. 
Stacking such layers in the growth direction, in Section~\ref{sec:3} we construct a 3D system, exhibiting a nodal line, 
which, in a close analogy to the 2D film case, exists at a critical point, separating two distinct Weyl semimetal phases with opposite 
signs of the anomalous Hall conductivity. 
This construction allows us to make an explicit connection between a nodal line in a 3D system and a pair of massless Dirac 
fermions, separated in momentum space, in a 2D film. We also make a connection between the drumhead surface state of the
3D nodal line semimetal and the edge states of the double 2D Dirac fermion system. 
In Section~\ref{sec:4} we describe analogous physics in PT-symmetric nodal line semimetals. 
Based on the results of Sections~\ref{sec:2}, \ref{sec:3} and \ref{sec:4}, in Section~\ref{sec:5} we construct a field 
theory, which describes anomalous electromagnetic response of nodal line semimetals. 
We demonstrate that a necessary ingredient in this field theory is a vielbein field determinant, which encodes the chirality 
of the Weyl fermions. The vielbein determinant changes sign at the quantum phase transition, at which the chirality of the Weyl fermions changes sign, requiring the appearance of a nodal line. 
We conclude with a brief discussion of the experimental observability of the proposed phenomena in Section~\ref{sec:6}. 
\section{Thin TI film in an external field}
\label{sec:2}
We start by examining a simple, yet realistic, system, which will allow us to most clearly demonstrate the connection that exists between 2D Dirac fermions and 3D nodal lines. 
Let us consider a thin film of a 3D topological insulator material. 
Assuming the bulk material is a good insulator with a significant bandgap, the low-energy physics may be described by focusing on the 2D Dirac surface states only. 
The corresponding Hamiltonian, assuming a single 2D Dirac fermion per surface of the film at a time reversal invariant momentum (TRIM), which we 
will take to be at the $\Gamma$-point $\bk = 0$ of the BZ of the film for simplicity, is given by
\beq
\label{eq:4}
H_0 = \left[v_F (\hat z \times \bsigma) \cdot \bk + \frac{\lambda}{2}(k_+^3 + k_-^3) \sigma^z\right] \tau^z + \Delta \tau^x. 
\eeq
Here $\bsigma$ is the spin, $\btau$ describes the top and bottom surface pseudospin degree of freedom, and 
$\hat z$ is the growth direction of the film.  
The $\Delta \tau^x$ term describes tunneling between the top and the bottom surface, which is nonnegligible when the 
film is sufficiently thin, which means the thickness of a few unit cells in practice. 
The cubic in the crystal momentum term in the square brackets describes hexagonal warping of the surface states, present in 
Bi$_2$Te$_3$ and related 3D TI compounds.~\cite{Fu_warping}
This term is often omitted when discussing TI surface states, but it (or its analogs in materials with non-hexagonal crystal symmetry)
is always present and will play an important role in what follows. 
In particular this term breaks the continuous $C_{\infty}$ symmetry with respect to arbitrary rotations around the $z$-axis, which 
the linearized Hamiltonian of the TI film would possess, down to the physical $C_3$ symmetry. 

We now imagine doping the TI film with magnetic impurities, or placing it in proximity to a ferromagnetic insulator film, 
which induces a Zeeman spin-splitting term in Eq.~\eqref{eq:4}
\beq
\label{eq:5}
H = H_0 + \bb \cdot \bsigma, 
\eeq
where the direction of the vector $\bb$ is arbitrary and may be parametrized in the standard way using the polar and azimuthal angles 
as $\bb = b (\sin \theta \cos \phi, \sin \theta \sin \phi, \cos \theta)$. 
This sort of a system has been created experimentally, most notably in the context of realizing the 2D quantum anomalous Hall effect (QAHE).~\cite{Chang13,QAHE}
\begin{figure}[t]
\includegraphics[width=12cm]{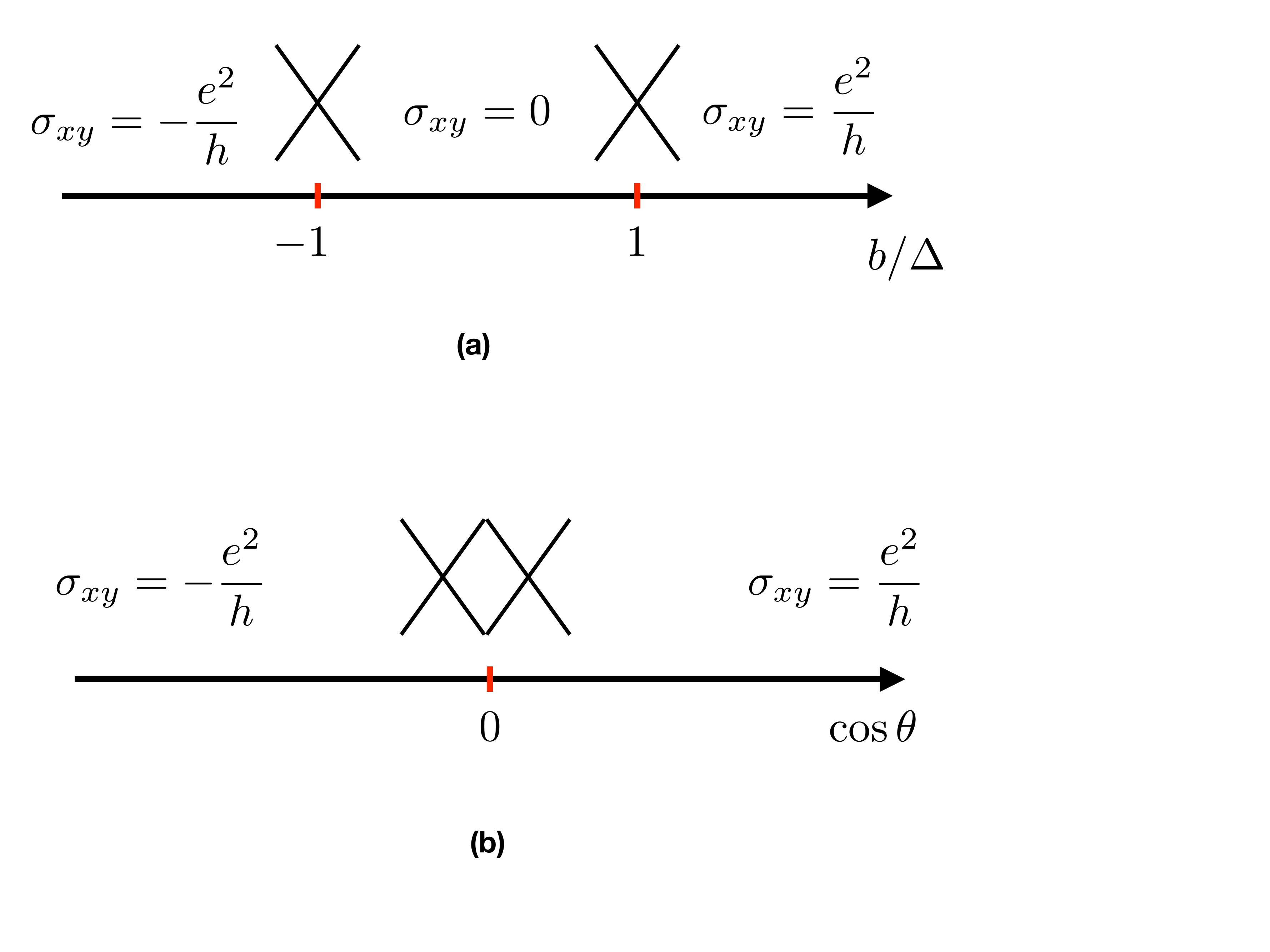}
\caption{(Color online) Phase diagram of a thin TI film in a spin-splitting field $\bb$. 
(a) The field is along the growth ($z$) direction of the film. The two QAHE phases with 
$\sigma_{xy} = \pm e^2/h$ are separated by an intermediate normal insulator phase with $\sigma_{xy} = 0$. 
Each of the two transitions is described by a single massless 2D Dirac fermion. (b) The field $\bb$ is rotated in the $xz$-plane while 
its magnitude $b > \Delta$ is held fixed. In this case, a direct transition between the two QAHE phases is possible. Two massless 2D Dirac fermions, separated in momentum space, must emerge at the critical point to produce a jump of $2e^2/h$ in the Hall conductivity.}
\label{fig:1}
\end{figure}
\begin{figure}[t]
\subfigure[]{
\includegraphics[width=7cm]{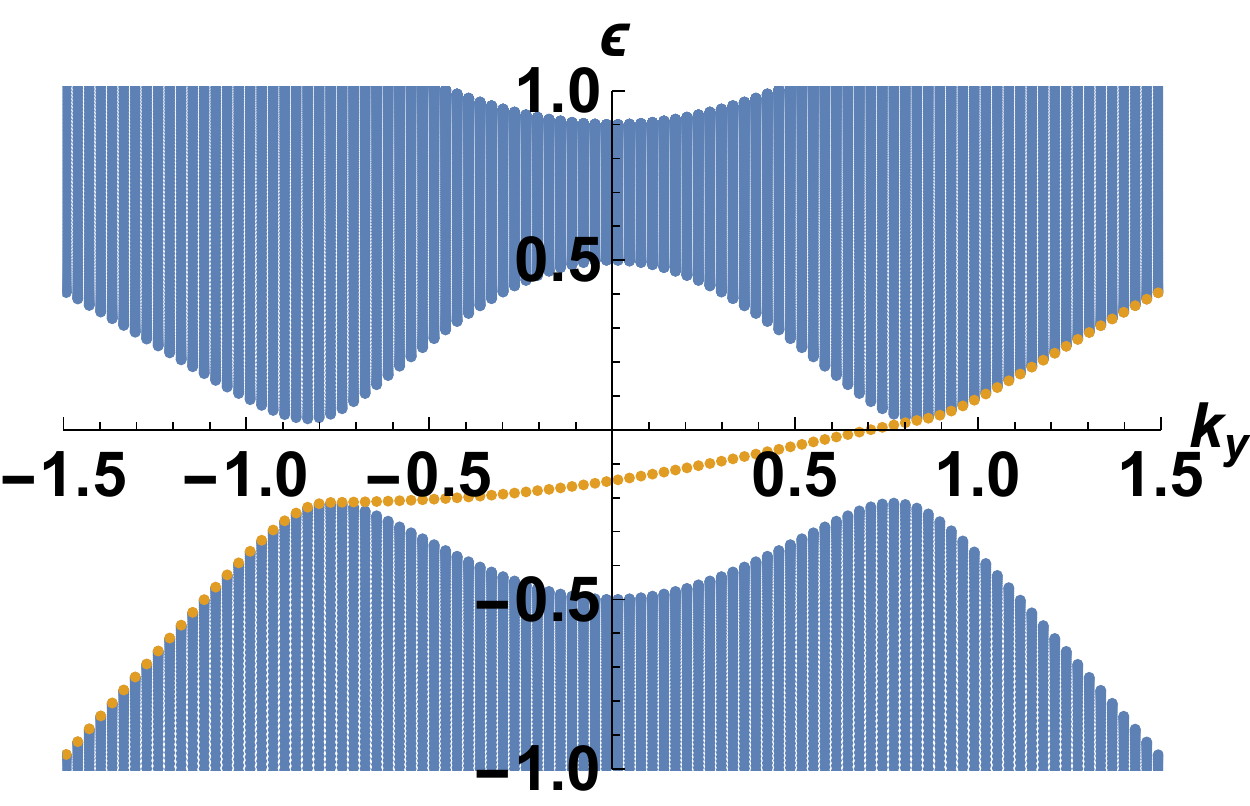}}
\subfigure[]{
\includegraphics[width=7cm]{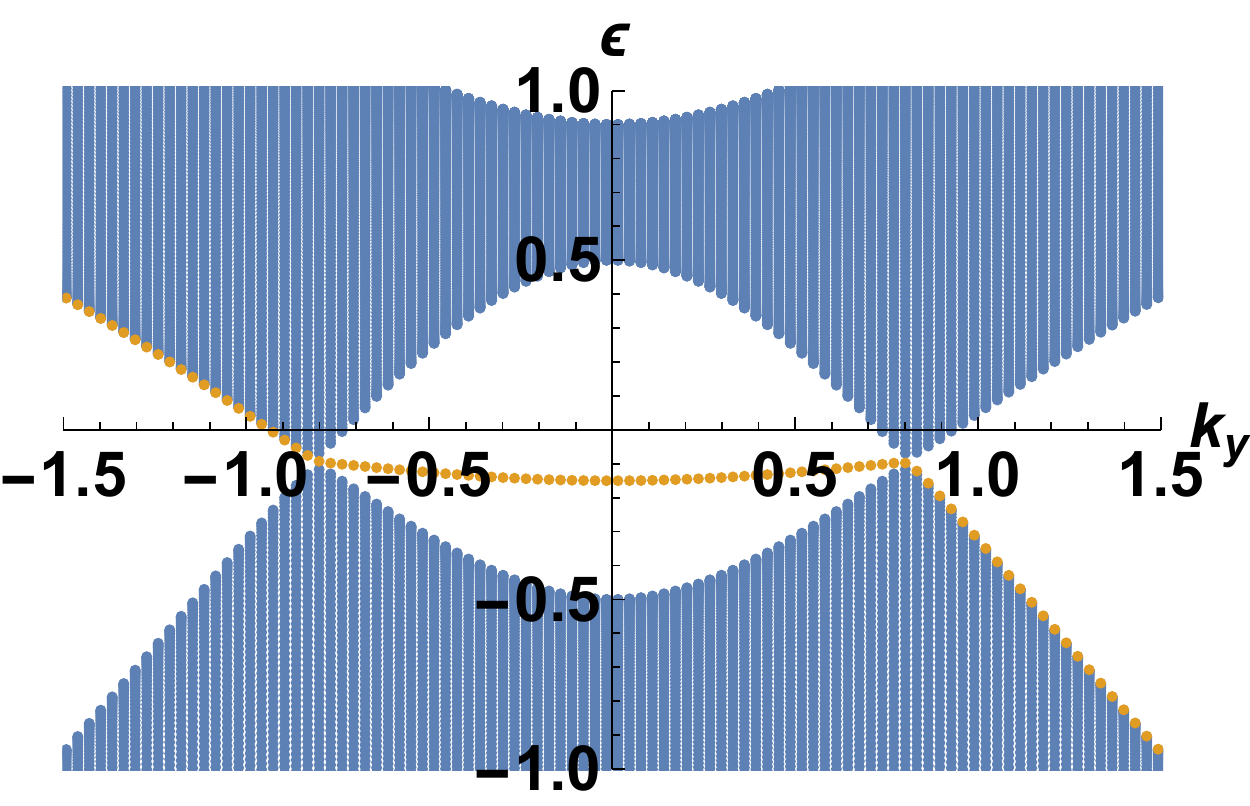}}
\subfigure[]{
\includegraphics[width=7cm]{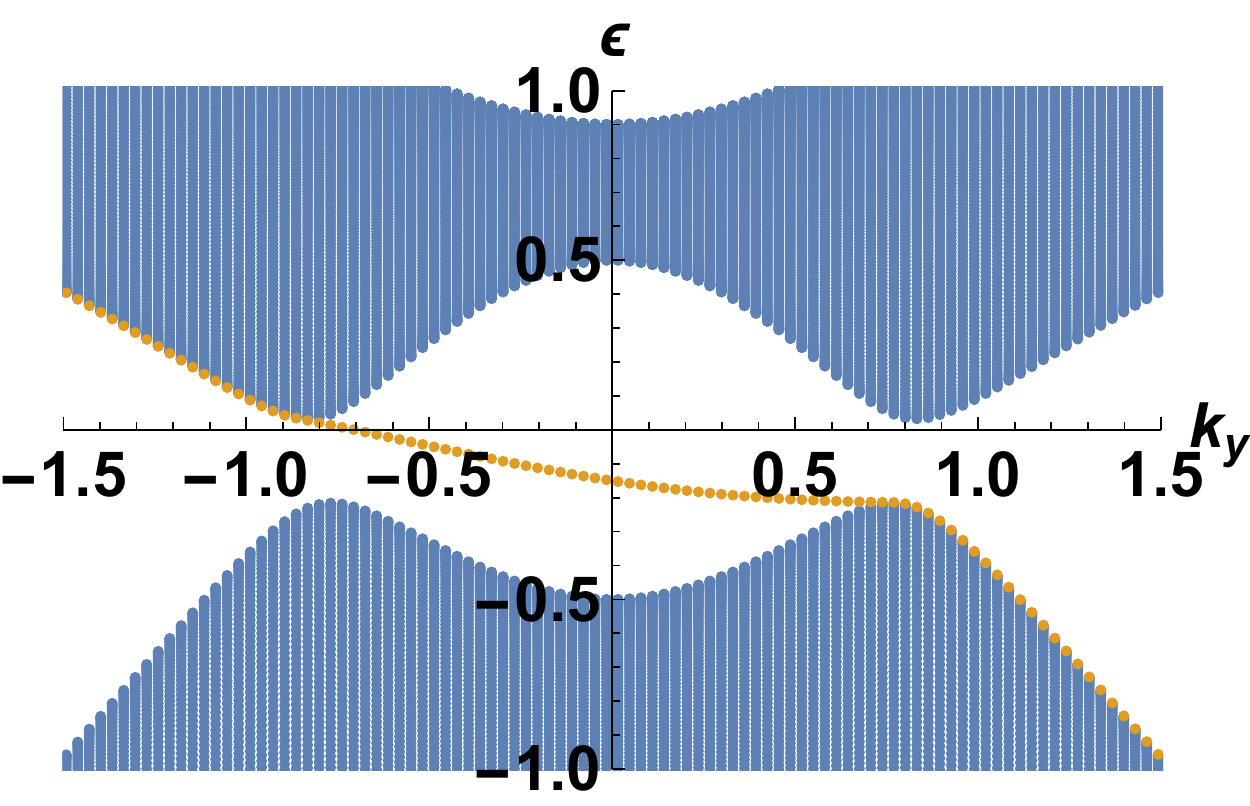}}
\caption{(Color online) Edge states of a TI thin film sample in the form of a strip. (a) Chiral edge state on one of the edges of the film
(the other edge state not shown for clarity) 
for $\theta < \pi/2$. (b) Edge state at the transition point $\theta = \pi/2$, which connects the two bulk Dirac points. (c) Edge state
with opposite chirality for $\theta > \pi/2$. We have included particle-hole symmetry breaking terms, omitted in Eq.~\eqref{eq:4}, for the sake of generality.}
\label{fig:2}
\end{figure}
We want to understand the phase diagram of this system as a function of the magnitude and the direction of the spin-splitting 
field $\bb$. 

Let us start from the case when $\bb = b \hat z$ is along the growth direction of the film. 
A similarity transformation 
\beq
\label{eq:6}
\sigma^{x,y} \rightarrow \tau^z \sigma^{x,y}, \,\, \tau^{x,y} \rightarrow \sigma^z \tau^{x,y},
\eeq
brings the Hamiltonian to the form (the warping term may be ignored here)
\beq
\label{eq:7}
H = v_F (\hat z \times \bsigma) \cdot \bk + (b + \Delta \tau^x) \sigma^z. 
\eeq
This in turn may be brought to the following block-diagonal form by diagonalizing the $\tau^x$ matrix
\beq
\label{eq:8}
H_r = v_F (\hat z \times \bsigma) \cdot \bk + m_r \sigma^z, 
\eeq
where $r = \pm$ are the two eigenvalues of $\tau^x$ and $m_r = b + r \Delta$. 
Each of the two $2 \times 2$ Hamiltonians $H_r$ describes a 2D Dirac fermion of ``mass" $m_r$. 
The $m_-$ ``mass" goes to zero and changes sign at $b = \Delta$, marking a quantum Hall plateau transition between 
a normal insulator state with $\sigma_{xy} = 0$ when $b < \Delta$ to a quantum anomalous Hall insulator state with $\sigma_{xy} = e^2/h$ when $b > \Delta$. The critical point is a described by a massless 2D Dirac fermion, centered at the $\Gamma$-point (or any TRIM more generally) in the first BZ.~\cite{Ludwig94}

Now suppose we rotate $\bb$ away from the $z$-axis. 
Here the response of the system depends crucially on the azimuthal angle $\phi$, i.e. it depends on the plane in which the 
field is rotated away from the $z$-axis. This angular dependence exists due to the presence of the hexagonal warping term. 
Let us start by rotating the field in the $xz$-plane, which corresponds to $\phi = 0$. 
We note that when the field is rotated all the way to the $x$-axis, which corresponds to $\theta = \pi/2$, the Hamiltonian of the 
film Eq.~\eqref{eq:5} possesses the following symmetry
\beq
\label{eq:9}
\sigma^x H(-k_x, k_y) \sigma^x = H(k_x, k_y). 
\eeq
Physically this symmetry is simply the mirror reflection symmetry with respect to the $yz$-plane, which exists even in the 
presence of the hexagonal warping term since $k_+^3 + k_-^3 = 2 k_x(k_x^2 - 3 k_y^2)$. 
This symmetry has important consequences for how the system responds to the field $\bb$, when it is rotated 
in the $xz$-plane and is thus normal to the mirror reflection plane $k_x = 0$ when $\theta = \pi/2$.  

To see what happens we again block-diagonalize Eq.~\eqref{eq:5} by first rotating the spin quantization axis along the 
direction of $\bb$ and then performing the similarity transformation Eq.~\eqref{eq:6}. 
This brings the Hamiltonian to the form 
\beqa
\label{eq:10}
H_r&=&\left[v_F \cos \theta \, k_y - \frac{\lambda}{2}\sin \theta \, (k_+^3 + k_-^3)\right] \sigma^x - v_F \sigma^y k_x \nonumber \\
&+&m_r(\bk) \sigma^z, 
\eeqa
where 
\beq
\label{eq:11}
m_r(\bk) = b + r \sqrt{\left[v_F \sin \theta k_y + \frac{\lambda}{2} \cos \theta (k_+^3 + k_-^3)\right]^2 + \Delta^2}, 
\eeq
and $r = \pm$ as before. 
For all $\theta \neq \pi/2$ the spectrum of $H_r$ has a full gap. 
When $\theta = \pi/2$, however, there are two Dirac band-touching points on the $y$-axis, whose coordinates are given by the 
solution of the equation
\beq
\label{eq:12}
m_-(k_x = 0, k_y) = b - \sqrt{v_F^2 k_y^2 + \Delta^2} = 0, 
\eeq
which gives 
\beq
\label{eq:13}
k_y^{\pm} = \pm \frac{1}{v_F} \sqrt{b^2 - \Delta^2}, 
\eeq
assuming $b > \Delta$.  

There are two different ways to understand this result. 
First, if we take both $\lambda$ and $\Delta $ to zero in Eq.~\eqref{eq:5}, it is clear that the 
field in the $x$-direction simply shifts the gapless top and bottom surface states of the TI film to different 
points on the $y$-axis with coordinates $k_y^{\pm} = \pm b/v_F$ in the BZ. 
When $\Delta > 0$, which produces a NI state in the absence of the spin-splitting field, a transition to a semimetallic state 
with two Dirac points happens at $b = \Delta$,~\cite{Zyuzin11} with the Dirac point locations given by Eq.~\eqref{eq:13}.
Crucially, even when the hexagonal warping term is included, mirror reflection symmetry with respect to the $yz$-plane forces it 
to vanish everywhere on the $y$-axis and protects the gaplessness of the Dirac points, even though time reversal 
symmetry is violated by the nonzero $\bb$ field. 

Another way to view the reappearance of the gapless Dirac points when 
$\bb$ is rotated along the $x$-direction, is revealed when one considers the behavior of the anomalous Hall conductivity of the 
TI film, $\sigma_{xy}$. 
As discussed above, $\sigma_{xy} = e^2/h$ when $\bb = b \hat z$ and $b > \Delta$. If the field is rotated to the negative $z$-direction, the sign of the Hall conductivity will change to $\sigma_{xy} = -e^2/h$. 
The transition between the two quantized values, as the field is rotated from $z$ to $-z$ direction, can in general happen at any 
value of the polar angle $\theta$. However, when the field is rotated in the $xz$-plane, the transition is forced to happen at 
$\theta = \pi/2$ by the mirror reflection symmetry with respect to the $yz$-plane that exists only at this angle. 
This is because this mirror symmetry also forces $\sigma_{xy}$ to vanish at $\theta = \pi/2$.~\cite{Liu_QAHE}
Indeed, if $j_x = \sigma_{xy} E_y$, mirror reflection with respect to the $yz$-plane changes $j_x \rightarrow - j_x$, while 
$E_y$ does not change. Thus $\sigma_{xy} = 0$ in this case. 
The reason that two gapless Dirac points must appear at the transition is that $\sigma_{xy}$ changes by $2e^2/h$ 
as $\theta$ is rotated through $\pi/2$, each 2D Dirac fermion contributing a quantum of Hall conductance $e^2/h$ when its 
``mass" changes sign. 
The phase diagram of this system as a function of the angle $\theta$ is shown in Fig.~\ref{fig:1}. 

Let us now demonstrate explicitly that when the field is rotated in a different plane, which is not perpendicular to a 
mirror plane, the transition from $\sigma_{xy} = e^2/h$ to $-e^2/h$ happens at a nonuniversal angle, which depends on details 
of the Hamiltonian of the TI film. 
Suppose we rotate $\bb$ in the $yz$-plane now, which corresponds to $\phi = \pi/2$. 
The block-diagonalized Hamiltonian in this case is given by
\beqa
\label{eq:14}
H_r&=&\left[-v_F \cos \theta \, k_x - \frac{\lambda}{2}\sin \theta \, (k_+^3 + k_-^3)\right] \sigma^x - v_F \sigma^y k_y \nonumber \\
&+&m_r(\bk) \sigma^z, 
\eeqa
where 
\beq
\label{eq:15}
m_r(\bk) = b + r \sqrt{\left[-v_F \sin \theta k_x + \frac{\lambda}{2} \cos \theta (k_+^3 + k_-^3)\right]^2 + \Delta^2}, 
\eeq
The gap in this case may close only on the $x$-axis at two points with the coordinates
\beq
\label{eq:16}
k^{\pm}_x = \pm \sqrt{\frac{-v_F \cot \theta_c}{\lambda}}, 
\eeq
which means that $\theta_c > \pi/2$ when $\lambda  > 0$. 
The critical polar angle $\theta_c$, at which the plateau transition from $\sigma_{xy} = e^2/h$ to $\sigma_{xy} = -e^2/h$ happens,
satisfies the following equation
\beq
\label{eq:17}
\cot^3 \theta_c + \cot \theta_c + \frac{\lambda}{v_F^3} (b^2 - \Delta^2) = 0. 
\eeq
If $\lambda (b^2 - \Delta^2)/v_F^3 \ll 1$, we may ignore $\cot^3 \theta_c$ above and obtain
\beq
\label{eq:18}
\theta_c = \textrm{arccot} \left[- \frac{\lambda (b^2 - \Delta^2)}{v_F^3}\right]. 
\eeq
In this case the locations of the Dirac points on the $x$-axis are given by
\beq
\label{eq:19}
k_x^{\pm} = \pm \frac{1}{v_F} \sqrt{b^2 - \Delta^2}. 
\eeq

Now let us go back to the case of the field rotated in the $xz$-plane (the mirror symmetry that exists in this case will play a crucial role later, when we discuss the nodal line semimetals) and establish a connection between the anomalous response
and the edge states, along the lines of the discussion in the Introduction. 
As established above, the anomalous Hall conductivity as a function of the polar angle $\theta$ has the following form in this case
\beq
\label{eq:20}
\sigma_{xy} = \frac{e^2}{h} \textrm{sign} (\cos \theta). 
\eeq
This singular dependence on the angle may be viewed as being a consequence of the parity anomaly~\cite{Semenoff84,Haldane88} of the two massless 2D Dirac fermions, which appear at $\theta = \pi/2$ in the presence of the mirror reflection symmetry.
The quantized $\sigma_{xy}$ is associated with chiral 1D edge states, whose chirality determines the sign of 
$\sigma_{xy}$. This means that at $\theta = \pi/2$ the edge states must switch their chirality. 
This implies that the critical state with two Dirac points separated in momentum space, realized at $\theta = \pi/2$, must itself have edge states, which connect the two Dirac points, as shown in Fig.~\ref{fig:2}. 
\begin{figure}[t]
\includegraphics[width=9cm]{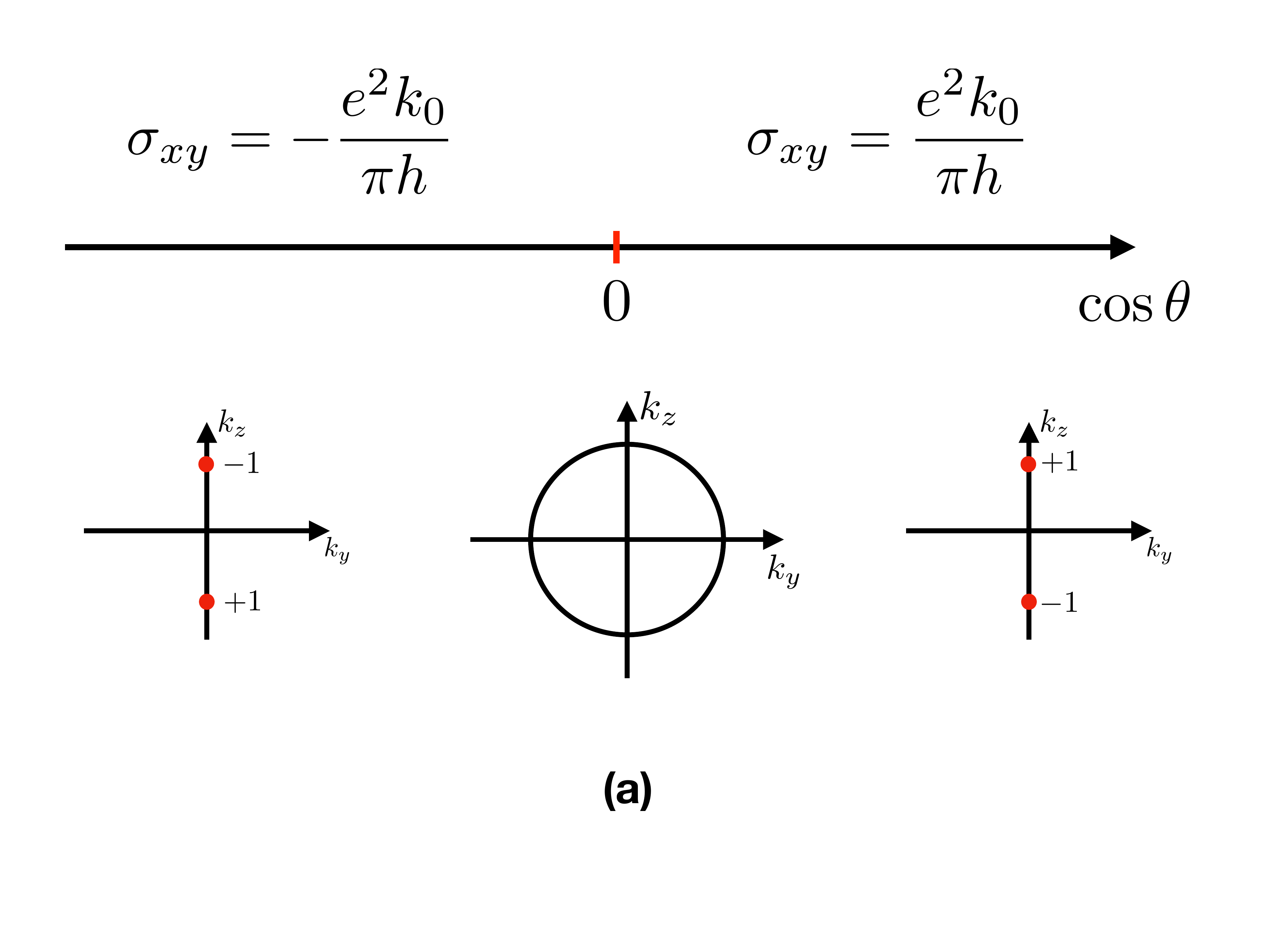}\\
\includegraphics[width=9cm]{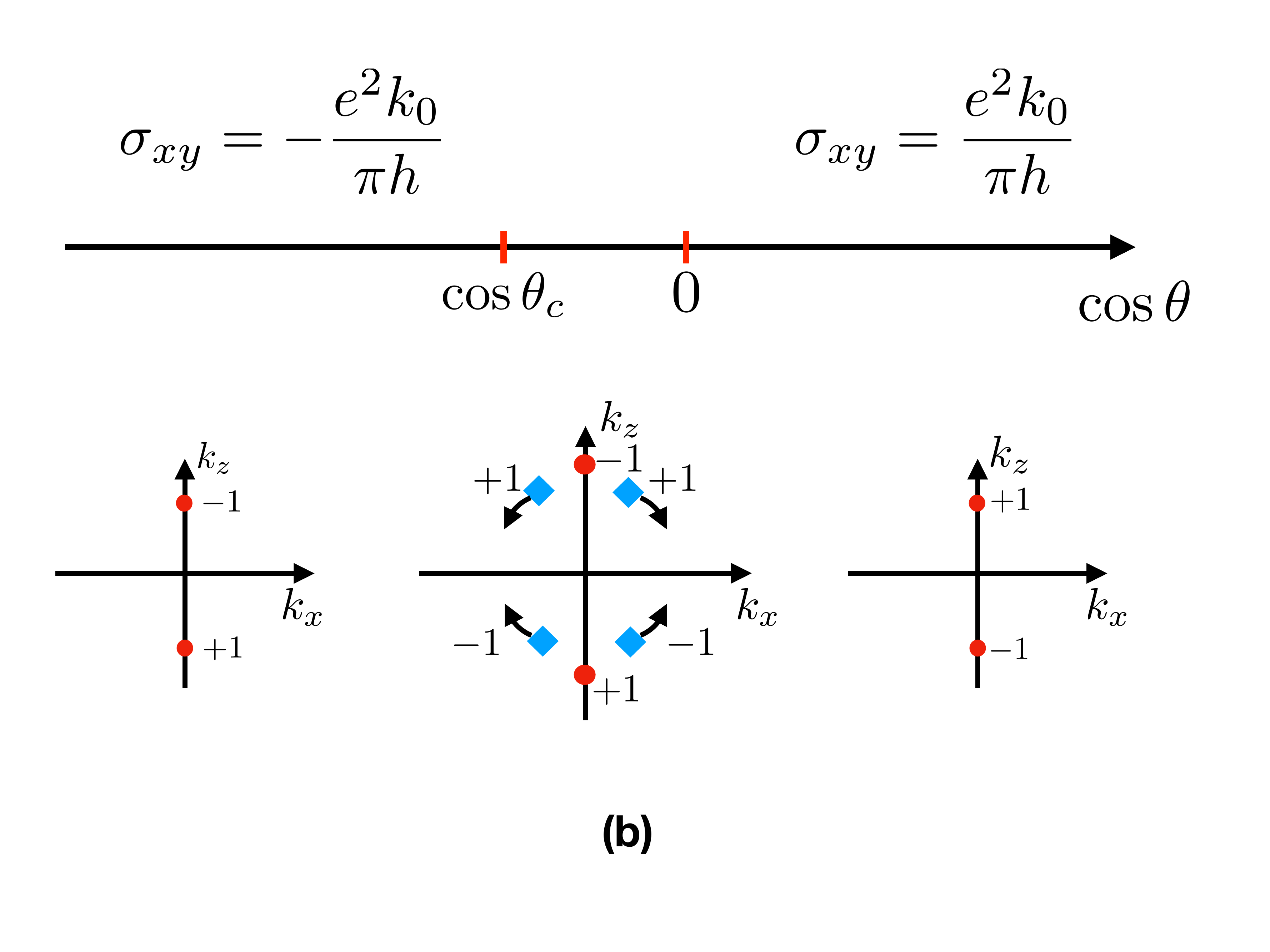}
\caption{(Color online) Phase diagram of the TI-NI multilayer as a function of the direction of the spin-splitting field $\bb$. 
(a) The field is rotated in the $xz$-plane, so that when $\theta = \pi/2$ the field $\bb$ is normal to the mirror plane $k_x = 0$. 
In this case there is a direct transition between $\sigma_{xy} = \pm e^2 k_0/ \pi h$ with a nodal line at the critical point. 
(b) The field is rotated in the $yz$-plane, so it is never normal to a mirror plane. In this case there is no direct transition 
between $\sigma_{xy} = \pm e^2 k_0/ \pi h$ and instead of a nodal line there is an intermediate phase with two extra Weyl node pairs. 
$\sigma_{xy}$ changes smoothly within this phase, as given by Eq.~\eqref{eq:33}.}
\label{fig:3}
\end{figure}

These edge states may naively appear similar to the Fermi arcs of Weyl semimetals. 
However, there is an important difference between them. 
The Fermi arcs of 3D Weyl semimetals are chiral and connect 
the conduction and valence bands, apart from connecting the Weyl nodes. This means that both their location within the gap 
and their connection to the Weyl nodes are topologically protected. 
In contrast, the 1D edge states that connect the 2D Dirac points in Fig.~\ref{fig:2}b are not chiral and may be pushed out of the gap and made to merge with the bulk states by a sufficiently strong perturbation. 
The only topologically protected property they have is that they always connect to the bulk Dirac points.

\section{From 2D Dirac points to 3D nodal line}
\label{sec:3}
We now extend the physics, discussed in Section~\ref{sec:2}, to 3D, by making a multilayer stack of 
thin TI films.~\cite{Burkov11-1}
The Hamiltonian of the stack is given by
\beqa
\label{eq:21}
H&=&\left[v_F (\hat z \times \bsigma) \cdot \bk + \frac{\lambda}{2} (k_+^3 + k_-^3) \sigma^z \right] \tau^z \nonumber \\
&+&(\Delta_S + \Delta_D \cos k_z) \tau^x - \Delta_D \sin k_z \tau^y + \bb \cdot \bsigma,
\eeqa
where $\Delta_S$ and $\Delta_D$ describe tunneling between the 2D Dirac surface states within the same (S) or neighboring (D) 
TI layers and we will take the period of the heterostructure in the growth direction to be the unit of length. 
As before, let us assume that we may rotate the spin-splitting field $\bb$ in two planes: $xz$-plane, corresponding to $\phi = 0$ 
and $yz$-plane, corresponding to $\phi = \pi/2$. The difference between the two planes is that $yz$ is a mirror-symmetric plane, while 
$xz$ is not, but is normal to a mirror symmetric plane instead. 
\subsection{Mirror symmetric case}
Let us start with the case of the field rotated in the $xz$-plane. In the case, the multilayer will have mirror symmetry when the field 
is along the $x$-axis and is thus perpendicular to a mirror plane ($yz$-plane). 
The Hamiltonian is block-diagonalized by exactly the same transformations as in Section~\ref{sec:2} and we obtain
\beqa
\label{eq:22}
H_r&=&\left[v_F \cos \theta \, k_y - \frac{\lambda}{2}\sin \theta \, (k_+^3 + k_-^3)\right] \sigma^x - v_F \sigma^y k_x \nonumber \\
&+&m_r(\bk) \sigma^z, 
\eeqa
where 
\beq
\label{eq:23}
m_r(\bk) = b + r \sqrt{\left[v_F \sin \theta k_y + \frac{\lambda}{2} \cos \theta (k_+^3 + k_-^3)\right]^2 + \Delta^2(k_z)}, 
\eeq
and 
\beq
\label{eq:24}
\Delta(k_z) = \sqrt{\Delta_S^2 + \Delta_D^2 + 2 \Delta_S \Delta_D \cos k_z}. 
\eeq
This is identical to the result obtained before in the single TI film case, Eqs.~\eqref{eq:10} and \eqref{eq:11}, 
except the constant tunneling amplitude $\Delta$ is replaced by the function $\Delta(k_z)$. 
Thus we may use the results of the previous section by treating every value of $k_z$ as parametrizing an effective 2D TI film system 
with the tunneling amplitude $\Delta(k_z)$. 
In particular, for all $\theta \neq \pi/2$, every value of $k_z$ corresponds to a gapped 2D insulator with either 
$\sigma_{xy}(k_z) = e^2/h$ 
or $\sigma_{xy}(k_z) = 0$, except when $m_-(k_x = 0, k_y=0, k_z) = 0$, which corresponds to a critical point between the two
insulators. The solution of this equation gives the locations of two Weyl points on the $z$-axis, given by
\beq
\label{eq:25}
k_z^{\pm} = \pi \pm k_0 = \pi \pm \textrm{arccos}\left(\frac{\Delta_S^2 + \Delta_D^2 - b^2}{2 \Delta_S \Delta_D}\right). 
\eeq
This corresponds to the anomalous Hall conductivity, given by 
\beq
\label{eq:26}
\sigma_{xy} = \int_{k_z^-}^{k_z^+} \frac{d k_z}{2 \pi} \sigma_{xy}(k_z) = \frac{e^2 k_0}{\pi h}, 
\eeq
which is obtained by summing contributions of effective 2D systems for every value of $k_z$ in between the Weyl node 
locations.~\cite{Burkov11-1}

When $\theta =  \pi/2$, the field is normal to a mirror plane $k_x = 0$ and at every $k_z$ we get two 2D Dirac points, 
signifying a transition, at which the sign of $\sigma_{xy}(k_z)$ changes. 
The locations of the Dirac points at every $k_z$ satisfy the equation
\beq
\label{eq:27}
\sqrt{v_F^2 k_y^2 + \Delta^2(k_z)} = b. 
\eeq
This equation clearly defines a closed curve in the $yz$-plane in momentum space (the mirror plane). 
This closed curve is a nodal line, at which the two bands, corresponding to the eigenvalue $r = -1$ touch. 
This line exists because of the mirror symmetry with respect to the $yz$-plane when $\bb$ is perpendicular 
to this plane (i.e. is along the $x$-axis). 
The nodal line may thus be regarded as a critical state, separating two Weyl semimetals with opposite sign of the anomalous Hall 
conductivity. 
\begin{figure}[t]
\includegraphics[width=9cm]{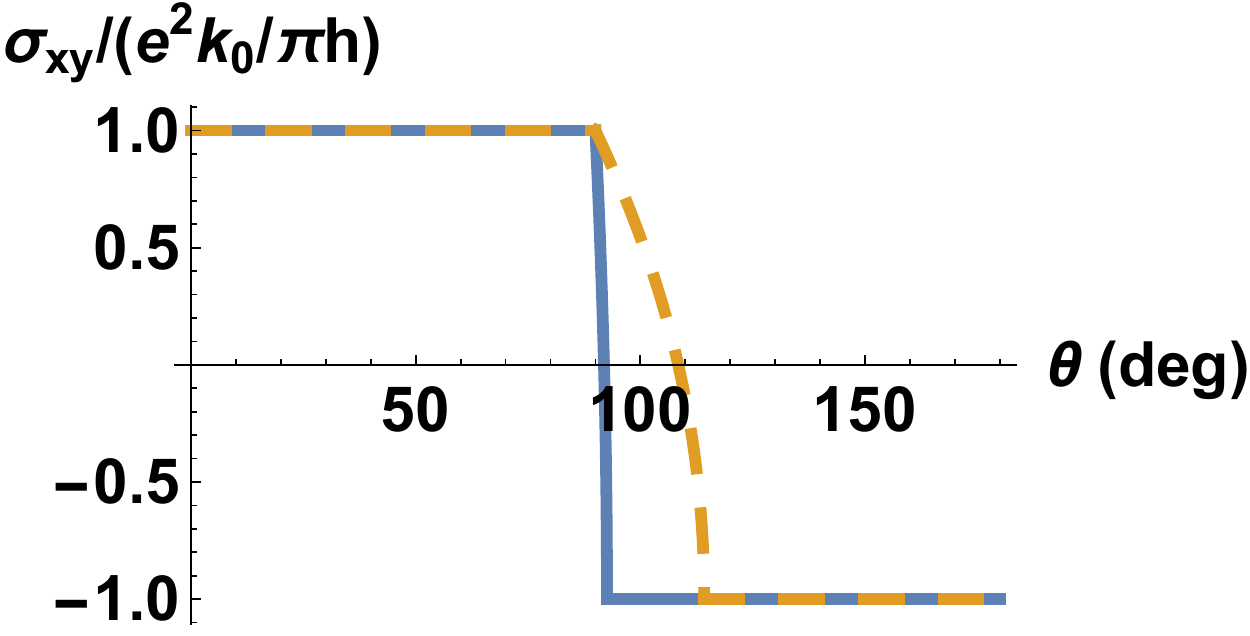}
\caption{(Color online) Anomalous Hall conductivity of the TI-NI multilayer system as a function of the polar angle $\theta$ when the spin-splitting field $\bb$ is rotated in the $yz$ plane and is never normal to a mirror plane for two different values of the parameter 
$\delta$, which determines the strength of the mirror symmetry violation by the hexagonal warping term: $\delta = 0.05$ (solid line) and $\delta = 0.5$ (dashed line).}
\label{fig:4}
\end{figure}
\subsection{Case without mirror symmetry}
Now let us see what happens when $\bb$ is rotated in a mirror ($yz$) plane, so that the mirror symmetry is never present, since it 
is always broken by the field.  
In this case we have
\beqa
\label{eq:28}
H_r&=&- \left[v_F \cos \theta \, k_x + \frac{\lambda}{2}\sin \theta \, (k_+^3 + k_-^3)\right] \sigma^x - v_F \sigma^y k_y \nonumber \\
&+&m_r(\bk) \sigma^z, 
\eeqa
where 
\beqa
\label{eq:29}
m_r(\bk)&=&b \nonumber \\
&+&r \sqrt{\left[-v_F \sin \theta k_x + \frac{\lambda}{2} \cos \theta (k_+^3 + k_-^3)\right]^2 + \Delta^2(k_z)}. \nonumber \\ 
\eeqa
The gap in this case may only close on the $k_y = 0$ plane, when the following equations are satisfied
\beqa
\label{eq:30}
&&k_x^2 = -\frac{v_F}{\lambda} \cot \theta, \nonumber \\
&&\cot^3 \theta + \cot \theta + \frac{\lambda}{v_F^3} [b^2 - \Delta^2(k_z)] = 0, 
\eeqa
i.e. the critical angle in this case is actually a function of $k_z$. This implies that the nodal line does not exist. 
The critical point, at which the sign of the 3D Hall conductivity Eq.~\eqref{eq:26} flips, also does not exist and is replaced 
by a smooth crossover, which is possible since a 3D Hall conductivity, unlike a 2D one, is not quantized. 

Specifically, the way the crossover occurs is as follows. 
The first of Eq.~\eqref{eq:30} clearly implies that $\theta_c \geq \pi/2$. 
Then both equations first acquire a real solution when $\theta = \pi/2$, $k_x = k_y = 0$ and $b = \Delta(k_z)$. 
This corresponds simply to the locations of the two Weyl points on the $z$-axis, $k_z^{\pm}$ of Eq.~\eqref{eq:25}. 
At this moment two pairs of extra Weyl points appear at these locations, which carry a topological charge, equal
to the topological charge of the original Weyl points. 
The topological charge of the original Weyl points themselves changes sign at this moment, so that the total charge at each $k_z^{\pm}$ is still the same. 
As $\theta$ increases past $\pi/2$, the extra two pairs of Weyl nodes shift away from the $z$-axis and move towards the $x$-axis
(see Fig.~\ref{fig:3}), 
mutually annihilating at a critical angle, given by the solution of the equation
\beq
\label{eq:31}
\cot^3 \theta_c + \cot \theta_c + \frac{\lambda}{v_F^3} [b^2 - \Delta^2(\pi)] = 0, 
\eeq
which, taking the last term in Eq.~\eqref{eq:31} to be small, so that the $\cot^3 \theta_c$ term can be neglected, gives
\beq
\label{eq:32}
\theta_c = \pi - \textrm{arccot}(\delta), 
\eeq
where 
\beq
\label{eq:32.5}
\delta = \frac{\lambda[b^2 - (\Delta_S - \Delta_D)^2]}{v_F^3}, 
\eeq
is a parameter that determines the strength of the mirror symmetry violation by the hexagonal warping term. 
Namely, the anomalous Hall conductivity changes smoothly (see Fig.~\ref{fig:4}) from $e^2 k_0/\pi h$ to $-e^2 k_0/\pi h$ in the interval 
$\pi/2 < \theta < \theta_c$, whose width is determined by the parameter $\delta$:
\begin{widetext}
\beqa
\label{eq:33}
\sigma_{xy} = \frac{e^2 k_0}{\pi h}\left\{
\begin{array}{c} 
1,\,\, 0 \leq \theta \leq \pi/2  \\ 
\frac{2}{k_0} \textrm{arccos}\left(\frac{\Delta_S^2 + \Delta_D^2 - b^2 -v_F^3 \cot \theta/\lambda}{2 \Delta_S \Delta_D} \right) - 1,\,\,\pi/2 \leq \theta \leq \theta_c \\
-1,\,\, \theta_c \leq \theta \leq \pi  
\end{array}
\right.,
\eeqa
\end{widetext}
which is calculated as in Eq.~\eqref{eq:26}. 
\subsection{Drumhead surface states}
Finally, let us come back to the case of the field rotated in the $xz$-plane. 
It is well known that the nodal line state, realized in this system when the field is rotated 
along the $x$-axis, i.e. is normal to the $yz$ mirror plane, is associated with the drumhead surface 
states.~\cite{Burkov11-2}
These surface states fill the projection of the nodal line onto the surface BZ and are dispersionless if particle-hole 
symmetry violating terms are ignored. 
As explained above, a 3D nodal line may be thought of as consisting of pairs of massless 2D Dirac fermions, separated 
in momentum space. 
We demonstrated in Section~\ref{sec:2} that such pairs of massless Dirac fermions lead to 1D edge states, which connect them.
This follows directly from the fact that a pair of massless 2D Dirac fermions describes the critical point between two quantum anomalous Hall insulators with opposite sign of the Hall conductivity. 
This implies that the drumhead surface states of 3D nodal line semimetals may be regarded as families of such 1D edge 
states, parametrized by the second momentum component. 
This also implies that, alternatively, the drumhead surface states may be regarded as chiral Fermi arcs of one of the two Weyl 
semimetal phases, separated by the critical point, described by the nodal line, in the limit when the chirality vanishes and switches 
sign. 

By an exact analogy with the argument given at the end of Section~\ref{sec:2} we observe that the drumhead surface states, 
unlike the chiral Fermi arcs, are not fully topologically protected in the absence of the particle-hole symmetry.~\cite{Burkov11-2,Bardarson17,Murakami_nl}
They may be pushed out of the gap and made to merge with the bulk states, but the protected property that remains is that they always connect to the bulk nodal line.

\section{Nodal line in PT-symmetric semimetals}
\label{sec:4}
So far we have been discussing the case of Weyl nodal lines, that is nodal lines in a material with broken time reversal 
symmetry, when two nondegenerate bands touch along the line. 
In this section we will demonstrate that similar ideas are applicable to the case of parity and time-reversal (PT) symmetric materials
in the absence of the spin-orbit (SO) interactions. In this case all bands are doubly degenerate with respect to the spin and touching 
is between two pairs of doubly-degenerate bands. 

This case is sufficiently simple that it may be understood without reference to a specific model. 
We may start from the most general Hamiltonian of a time reversal and parity invariant system with four degress 
of freedom per unit cell~\cite{Fu-Kane}
\beq
\label{eq:34}
H_0 = d_0(\bk) + \sum_{a=1}^5 d_a(\bk) \Gamma^a, 
\eeq
where $\Gamma_a$ are five gamma-matrices, obeying the Clifford algebra, which are even under the product 
of parity and time reversal. 
Taking Pauli matrices $\bsigma$ to represent the spin degree of freedom, and the eigenvalues of $\tau^z$ of another 
set of Pauli matrices $\btau$ to represent two orbital states in the unit cell, related to each other by parity (parity operator is 
$P = \tau^x$), they are given by
\beqa
\label{eq:35}
&&\Gamma_1 = \tau^x,\,\, \Gamma^2 = \tau^y,\,\, \Gamma^3 = \tau^z \sigma^x, \nonumber \\
&&\Gamma^4 = \tau^z \sigma^y,\,\, \Gamma^5 = \tau^z \sigma^z. 
\eeqa
In the absence of the SO interactions, only $\Gamma_1$ and $\Gamma_2$ may be present in the Hamiltonian. 
Ignoring the term, proportional to the unit matrix, for simplicity, we obtain
\beq
\label{eq:36}
H_0 = d_1(\bk) \tau^x + d_2(\bk) \tau^y. 
\eeq
Parity and time reversal symmetry require $d_1(\bk) = d_1(-\bk)$ and $d_2(\bk) = -d_2(-\bk)$. 
The electronic structure, described by $H_0$, exhibits a nodal line when $d_1(\bk) = d_2(\bk) = 0$. 

Now suppose we want to break time reversal symmetry, but keep the parity symmetry. 
This can be accomplished by adding a spin-splitting term $\bb \cdot \bsigma$, but the effect of this 
term in the absence of the SO interactions will be trivial, simply lifting the spin degeneracy. 
The only way to break time reversal symmetry but keep parity without involving the spin is to add a term $d_3(\bk) \tau^z$ 
to the Hamiltonian, where $d_3(\bk) = -d_3(-\bk)$.~\cite{Murakami17}
Microscopically, this would arise from a magnetic flux within the unit cell of the crystal, leading to Aharonov-Bohm 
phases of the hopping amplitudes, as in the Haldane Chern insulator model.~\cite{Haldane88}
To see the effect of this term on the nodal line, let us specify the functions $d_{1,2,3}(\bk)$. 
Let us take
\beq
\label{eq:37}
H_0(\bk) = (\Delta^2 - k_x^2 - k_y^2) \tau^x + v_1 k_z \tau^y. 
\eeq
The energy eigenvalues are
\beq
\label{eq:38}
\epsilon_{\pm}(\bk) = \pm \sqrt{(\Delta^2 - k_x^2 - k_y^2)^2 + v_1^2 k_z^2}. 
\eeq
The two bands touch along a nodal line in the $xy$-plane, given by the equation 
\beq
\label{eq:39}
k_x^2 + k_y^2 = \Delta^2. 
\eeq
Now let us add the time reversal breaking perturbation
\beq
\label{eq:40}
H = (\Delta^2 - k_x^2 - k_y^2) \tau^x + v_1 k_z \tau^y + v_2 k_x \tau^z. 
\eeq
The band energies become
\beq
\label{eq:41}
\epsilon_{\pm}(\bk) = \pm \sqrt{(\Delta^2 - k_x^2 - k_y^2)^2 + v_1^2 k_z^2 + v_2^2 k_x^2}. 
\eeq
We see that the nodal line has been gapped out except at two Weyl points on the $y$-axis at $k_y^{\pm} = \pm \Delta$. 
Taking $v_1 > 0$, the chirality of the two Weyl points is given by
\beq
\label{eq:42}
C_{\pm} = \pm \textrm{sign}(v_2), 
\eeq
and thus interchanges as $v_2$ is tuned through zero. 
Thus the nodal line in PT-symmetric systems has the same meaning as in systems with SO interactions and broken 
time reversal symmetry: this is a critical state that separates two Weyl semimetal states with opposite signs of the anomalous 
Hall conductivity.

\section{Field theory of nodal line semimetals}
\label{sec:5}
We will now summarize the above analysis of nodal line semimetals in terms of a field theory, which expresses 
their anomalous response, in the spirit of Eq.~\eqref{eq:1}, describing the chiral anomaly of Weyl semimetals. 
What we have established thus far is that the nodal line may be thought of as a state arising at a critical point between two
Weyl semimetal states with interchanged chirality of the Weyl points.  
This means that in addition to the vector $q_{\mu}$, which acts as a chiral gauge field, determining the separation of two 
Weyl nodes with fixed opposite chirality in momentum space, we need to take into account the possibility of chirality 
of the Weyl nodes changing sign, without changing their momentum-space location. 
This may be expressed with the help of the vielbein fields $e^{\mu}_a$, which encode both the effective metric and the chirality, characterizing the Weyl points.
A low-energy Hamiltonian of two Weyl nodes, separated in momentum space, may then be written as
\beq
\label{eq:43}
H = \gamma^0 \gamma^a \be_a \cdot (\bk - \bq \gamma^5), 
\eeq
which is a generalization of the ordinary massless Dirac Hamiltonian to a curved space-time with an arbitrary metric. 
The chirality-changing transition, with the appearance of a nodal line at the critical point, may be described as one of the 
vielbein vectors $\be_a$ flipping its direction and vanishing at the critical point.~\cite{Volovik17}
For example, in the case of the magnetized TI-NI multilayer, described by Eq.~\eqref{eq:22}, we may take 
\beq
\label{eq:44}
\be_2 = v_F \cos \theta\, \hat y, 
\eeq
which vanishes and changes direction at $\theta = \pi/2$. 
In the case of a PT-symmetric semimetal, described by Eq.~\eqref{eq:40}, we may take
\beq
\label{eq:45}
\be_1 = v_2 \hat x, 
\eeq
which again vanishes and changes direction when $v_2 = 0$. 
Defining chirality as 
\beq
\label{eq:46}
C_{\pm} = \pm \textrm{sign} [\be_1 \cdot (\be_2 \times \be_3)] = \pm \textrm{sign}[\textrm{det}(e)],
\eeq
where $\textrm{det}(e)$ is the determinant of the matrix $e^i_a$, 
we may then write down the following action, which describes the anomalous response of this system
\beq
\label{eq:47}
S = - \frac{e^2 q}{4 \pi^2} \textrm{sign}[\textrm{det}(e)] \int d t \, \, d^3 r\, \epsilon^{z \nu \alpha \beta} A_{\nu} \partial_{\alpha} A_{\beta}, 
\eeq
where we have taken the (fixed) direction of the vector $\bq$, which determines the separation between the Weyl nodes, 
to be the $z$-direction. 
This action bears a strong resemblance to the Chern-Simons action 
\beq
\label{eq:48}
S = - \frac{e^2}{4 \pi} \textrm{sign}(m) \int dt\, d^2 r \, \epsilon^{\nu \alpha \beta} A_{\nu} \partial_{\alpha} A_{\beta}, 
\eeq
which expresses the parity anomaly of a 2D Dirac fermion of mass $m$. 
This is not unexpected of course, given the connection between 3D nodal line and 2D Dirac fermions, which was established 
in Sections~\ref{sec:2} and \ref{sec:3}. 

The information, contained in Eq.~\eqref{eq:48} may be stated as follows: this equation is telling us that a massless 2D Dirac fermion describes a direct transition between nsulators with $\sigma_{xy} = 0$ and $\sigma_{xy} = e^2/h$.  
Analogously, Eq.~\eqref{eq:47} is telling us that the nodal line, which appears when one of the vielbein vectors $\be_i$ vanishes, describes a direct transition between two Weyl semimetal states with $\sigma_{xy} = \pm 2 e^2 q/ \pi h$.

\section{Discussion and conclusions}
\label{sec:6}
In this paper we have demonstrated that the anomalous response of 3D nodal line semimetals is closely 
related to the parity anomaly of 2D Dirac fermions. 
The role of the mass of a 2D Dirac fermion, whose sign enters into the topological Chern-Simons action for the 
electromagnetic field, when the fermions are integrated out, is played by the determinant of the vielbein matrix $e^i_a$. 

Perhaps the most straightforward way to observe these phenomena, given the currently available materials, is to 
look at magnetic response of type-II Dirac semimetals.~\cite{Burkov18-1}
Type-II Dirac semimetals possess one or several symmetry-related Dirac points at TRIM in the 
first BZ. 
Material realizations include TlBi(S$_{1-x}$ Se$_x$)$_2$,~\cite{Ando11}
(Bi$_{1-x}$In$_x$)$_2$Se$_3$,~\cite{Brahlek12} and ZrTe$_5$.~\cite{Li_anomaly,Chen15}
As was demonstrated in Ref.~\onlinecite{Burkov18-1}, magnetic response of a type-II Dirac point is always strongly 
anisotropic: while one of the Zeeman field components acts as a chiral gauge field, splitting the Dirac point into two Weyl points, 
the other two components instead create nodal lines. 
The problem then maps exactly onto the magnetic multilayer system, described in Section~\ref{sec:3} and exactly the same 
conclusions follow. 
The anomaly may then be detected as a step-function-like singularity of the anomalous Hall conductivity of the Dirac semimetal 
in the presence of an applied magnetic field, as the field is rotated. The anomalous Hall conductivity in this case is defined as part of the Hall conductivity, which arises from the Zeeman response. It may be obtained by subtracting off the linear high-field part of the 
Hall resistivity, as the anomalous Hall signal is usually isolated. 

Another possibility is the recently discovered magnetic Weyl semimetal Co$_3$Sn$_2$S$_2$.~\cite{Felser17}
In this material, six pairs of Weyl nodes arise out of nodal lines, gapped by the SO interactions, as revealed by the 
electronic structure calculations.~\cite{Felser17}
Thus Co$_3$Sn$_2$S$_2$ may naturally reside close to the phase transition at which the sign of the anomalous Hall conductivity changes, however a detailed investigation of how it responds to rotating the direction of magnetization is necessary to understand if this really is the case. 

In conclusion, we have presented a theory of anomalous response (quantum anomaly) in nodal line semimetals, which can be related 
to the existence of drumhead surface states in these systems. 
We have shown that both the surface states and the anomalous response are closely analogous to the parity anomaly 
of (2+1)-dimensional relativistic Dirac fermions, which in the condensed matter physics context is realized as the 2D QAHE. 
We have derived a field theory, describing the anomalous response of nodal semimetals, and shown that a crucial ingredient 
in this field theory is the sign of the determinant of the vielbein fields, describing both the 3D Weyl fermion chirality and the effective 
low-energy metric, which emerges in Weyl semimetals. This sign changes at a critical point at which one of the three vielbein vectors vanishes, leading to the emergence of a nodal line. 
\begin{acknowledgments}
Financial support was provided by NSERC of Canada. 
\end{acknowledgments}
\bibliography{references}
\end{document}